\documentclass[10pt,aps.prl,twocolumn,superscriptaddress]{revtex4-1} 
\usepackage{graphicx}
\usepackage{amsmath}
\usepackage{amssymb}
\usepackage{float}
\usepackage{subfig}
\usepackage{tabularx}
\usepackage[normalem]{ulem}
\usepackage{color}
\usepackage{cancel}
\usepackage{epstopdf}
\newcommand{\bibs}{d:/Dropbox/Dad/Mark/References/BibFile}
\newcommand{\enquote}[1]{``#1''}
\begin{document}
\title{Self-healing properties of 1-amino, 2,4-dibromo anthraquinone dye doped in PMMA polymer}
\author{Prabodh Dhakal}
\email{dprabodh@wsu.edu}
\author{Mark G. Kuzyk}
\email{mgk.wsu@gmail.com}
\affiliation{Department of Physics and Astronomy, Washington State University,
Pullman, WA 99164-2814}
\date{\today}

\begin{abstract}

We used fluorescence spectroscopic measurements as a probe to study the self-healing properties of anthraquinone derivative molecules doped in poly(methyl methacrylate) (PMMA). 2,4-dibromo anthraquinone dye doped in PMMA recovers after photodegradation. Its dynamics differs from other anthraquinone derivative molecules. This could be due to the relatively heavier bromine atom attached to one of the carbon atoms of the benzene ring. In this paper, we will discuss the self-healing properties of 2,4-dibromo anthraquinone doped in PMMA matrix. We also tested the correlated chromophore domain model (CCrDM) and have characterized the self-healing properties by determining the CCrDM parameters. We also estimated the self-absorption of fluorescence signal by the dye molecule without which the analysis of the self-recovery of the molecule would be incomplete.

\vspace{1em}
OCIS Codes:

\end{abstract}

\maketitle

\section{Introduction}

Researchers have studied the photodegradation and recovery of organic dyes\cite{Howell, Peng, Zhu-18, Brent, Zhang, Gonzalez, Y.Zhu} using different techniques ranging from a non-linear process such as two photons absorption\cite{Zhu-18, Y.Zhu} and amplified spontaneous emission (ASE)\cite{Howell, Brent, Embaye, Ramini}, to linear processes such as fluorescence\cite{Peng, Dhakal} and absorption spectroscopic measurements\cite{Embaye, P.Dhakal}. Each methods has advantages and drawbacks. For example, ASE is highly sensitive because it can differentiate between the undamaged and the damaged species, but it is extremely susceptible to laser drift and the nonlinear nature of the response makes it difficult to determine the population of the damaged molecules. As a result, the measured time constant is highly inconsistent. Laser drift over a long time period can distort the nature of the data and the mathematical function that describes the recovery kinetics. Similarly, in the case of fluorescence and absorption spectroscopy, the process is a linear, hence does not require a complex calibration. Furthermore, the amount of light absorbed is directly proportional to the population of the molecules in its path. In case of the absorption, both damaged and undamaged molecules contribute and one can not unentangle the spectra of the damaged and undamaged species\cite{Prabodh}. On the other hand fluorescence spectroscopy has been shown to be sensitive to only the undamaged species because the damaged species do not contribute significantly to the spectra measured.\cite{Prabodh}. Hence, we used fluorescence spectroscopy as a probe to study the self-healing properties of 2,4-dibromo anthraquinone molecule.

Of all the anthraquinone molecules that are shown to undergo self-recovery\cite{Prabodh}, the 2,4-dibromo anthraquinone molecule is of a special type because of the relatively heavier bromine atoms attached to the carbon atoms of the benzene rings. All the other dye molecules studied in the past have only nitrogen, oxygen or hydrogen on the benzene molecule. In an attempt to understand its recovery properties, we measured its decay and the recovery rate as a function of temperature and concentration.

The only disadvantage of fluorescence spectroscopy is self absorption because the sample by necessity has a thickness that exceeds the 1/e $ \times $ the absorption length\cite{Prabodh}. As a result, the fluorescence spectra partially overlaps with the absorption spectra. So the sample absorbs a part of the fluorescence signal on its way to the detector. Therefore, the true fluorescence spectra is altered, so the measured decay and recovery time constants can be inaccurate.

To account for this problem, we have estimated the self-absorption of the fluorescence signal. Self-absorption studied in the past\cite{Booth, Troger} does not apply to our geometry, so we have developed a customized model, which we have applied to the analysis of self-healing.

Finally, to understand the mechanism that governs self-recovery, we fit the data to the Correlated Chromophore Domain Model (CCrDM)\cite{Ramini, Anderson, Ben_Anderson}. Proposed mechanisms for self-healing include reorientational diffusion\cite{Abbas} and its absence\cite{ Embaye}, twisted intramolecular charge transfer\cite{ Dirk}, energy transfer between host polymer matrix and dye called exciton\cite{ Kasha} or between donor-acceptor sites\cite{Zhu, Nadumpara} and thermalization\cite{Fellows, Ishchenko}, but none of these mechanisms explain self-recovery data. Only the CCrDM fits into our data, suggesting domains are responsible.

\section{Experiment}

To study the self-healing properties of 2,4-dibromo anthraquinone molecules doped in PMMA polymer, we performed fluorescence spectroscopic measurements with the instruments diagrammed in Figure 1.

The sample under the study is a thin film prepared by the thermal-bulk-pressed method as follows. The monomer liquid of methyl methacrylate (MMA) is run through a column of densely packed quartz and aluminum powder to remove inhibitor and filter particulates. A given mass of 2,4-dibromo anthraquinone dye is dissolved into a volume of MMA liquid to get the desired concentration, which is immediately sonicated for 45-60 minutes to fully dissolve the dye homogenously into solution. A few drops of Butanethiol, a polymerization initiator, and tert-butyl peroxide, a chain transfer agent, are added to the dye-MMA solution to initiate the polymerization and control the polymer chain length. The solution is again sonicated for 10-15 mins to dissolve all the ingredients homogenously and passed through a GHP ACRODISC filter whose pore size is 0.2 $ \mu m $ to remove particulates or dye aggregates.

The solution thus obtained is then soaked at $ 95^{o}C $ for about 50 hours to polymerize it, then cooled, removed from the bottle and crushed to small chips. Some pieces of the chips are placed between two clean glass plates, squeezed for 90 mins to 110-120 psi of pressure at $ 150^{0}C $ in a custom built oven to form a thin film of 250-300$ \mu$m. Except for taking concentration dependent fluorescence measurements, all samples are prepared to a concentration of 3 grams of dye to 1 liter of MMA monomer. The thin film, thus prepared, is stored in a refrigerator.

As shown in the Figure 1, A pump laser beam of wavelength 488 nm from a Coherent Innova model 70C series ion laser falls on a pair of polarizers which controls the power of the pump laser beam to keep it at a constant average value of $ 3.46 \times 10^5 W/m^2 $. A clean glass plate splits a small portion of the pump laser beam to a Thorlabs S20MM model power meter which monitors the stability of the incident pump laser beam. Another portion of the laser beam is focused on the thin film sample by a convex lens. The sample is kept in an oven chamber whose temperature is controlled by a a resistive heater driven by a Micromega CN 77322-C2 unit, and the temperature is monitored with an Omega model CN-2010 Thermocouple. A manual shutter is used to launch the pump beam  onto the sample. The fluorescence signal produced by the sample is collected by a convex lens which directs the signal to a fiber coupled to an Ocean Optics model SD2000 spectrometer.

\begin{figure}[h!]
\begin{center}
\includegraphics{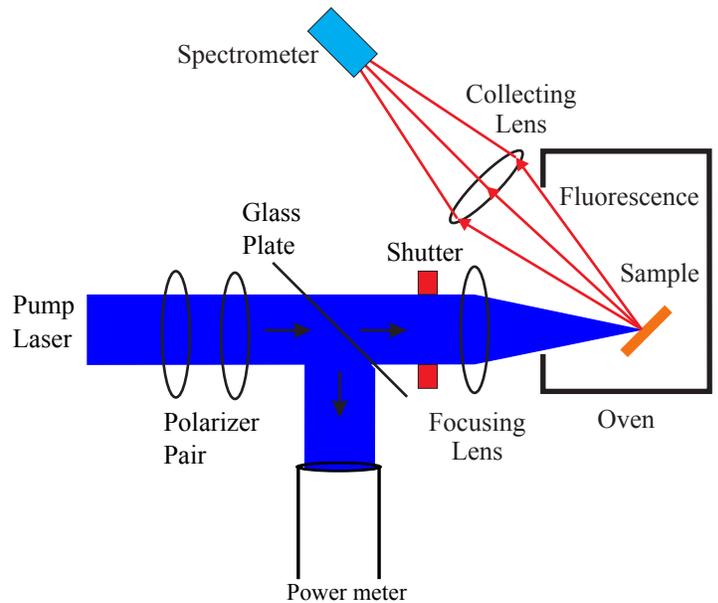}
\end{center}
\caption{fluorescence decay and recovery experiment.}
\label{fig:experiment}
\end{figure}

To estimate the degree of self-absorption by the sample and to calculate the CCrDM parameters, we performed a photochroism experiment, which is diagrammed in Figure 2. Except for the white light deuterium tungsten halogen source with a wavelength range of 215-2500 nm, all the other experimental components are the same as described in Figure 1. The white light passes through a filter to attenuate the intensity and a convex lens focusses it onto the sample. The spot at which the white light is focussed onto the sample coincides with that of the pump laser light and the diameter of the white probe beam must be smaller than the diameter of the pump laser beam to ensure that absorbance spectrum collected corresponds to the absorbance spectrum of the damaged species only. A manual shutter is used to block the white light during recovery and opened for 10-15 sec in every 30 minutes to take the absorbance spectra. Both fluorescence and absorbance spectra are collected by the same detector. 

\begin{figure}[h!]
\begin{center}
\includegraphics{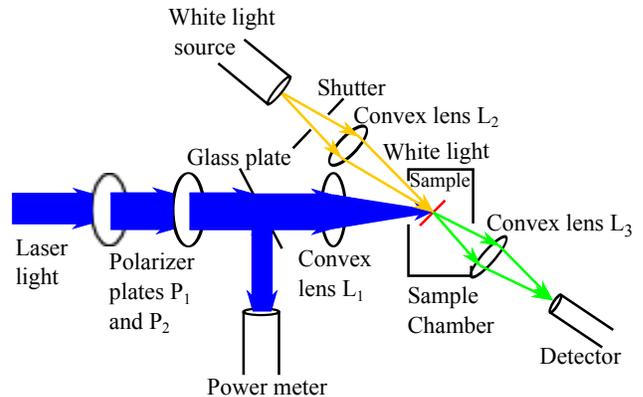}
\end{center}
\caption{Photochroism experiment. The pump beam continuously damages the sample while the probe beam measures the absorbance spectra}
\label{fig:photochroism}
\end{figure}

\section{Results and discussion}

\begin{figure}[h!]
\begin{center}
\includegraphics[scale=0.85]{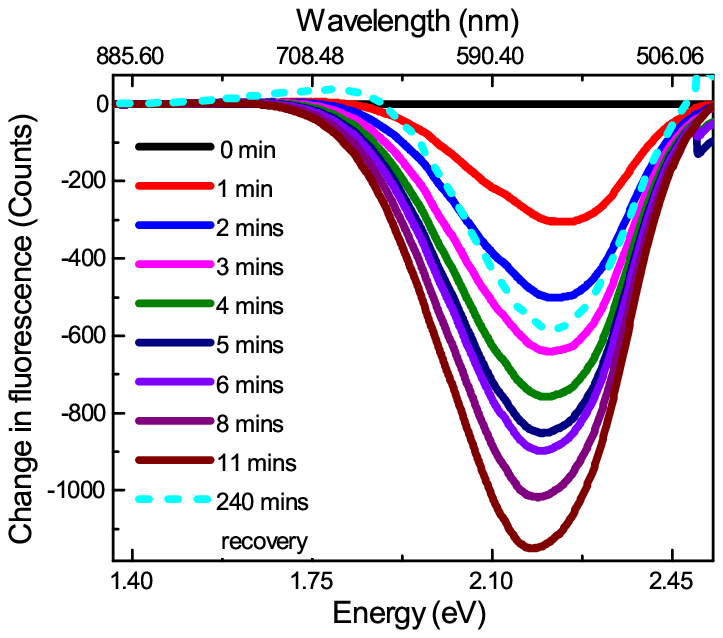} \\
\includegraphics[scale=0.85]{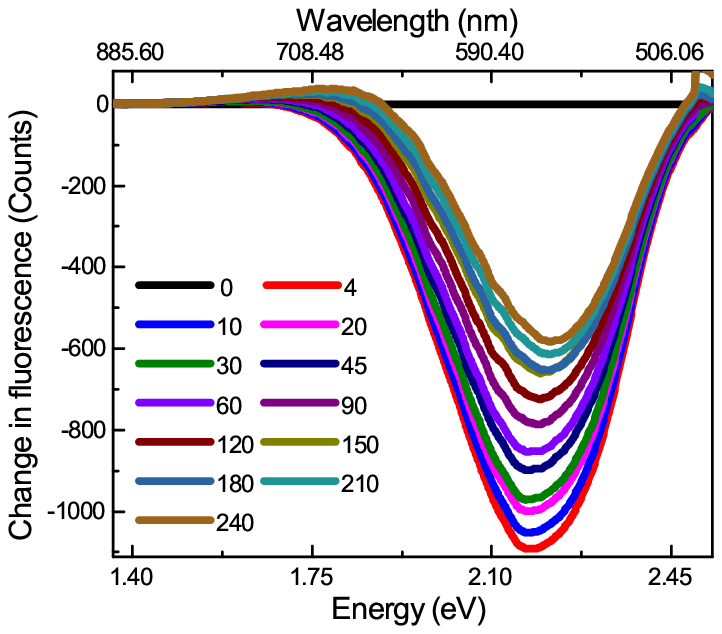}
\end{center}
\caption{Change in fluorescence spectra of 1-amino, 2,4-dibromo anthraquinone (top) during decay and (bottom) during recovery.}
\label{fig:flu_change_recovery}
\end{figure}

A representative change in fluorescence spectra of 1-amino, 2,4-dibromo anthraquinone dye during decay and recovery is shown in the Figure \ref{fig:flu_change_recovery}, which shows that there is no evidence or very weak evidence of an isosbestic point. We have discussed earlier\cite{Prabodh} that the absence of an isobestic point in the fluorescence spectrum suggest that only the fresh species of the anthraquinone dye molecules fluoresce while the damaged species either do not fluoresce or the fluorescence signal from the damaged species is  negligible compared to the fresh species. Hence, the fluorescence intensity measured is proportional to the population of the undamaged molecule. 

\begin{figure}[h!]
\begin{center}
\includegraphics[scale=0.85]{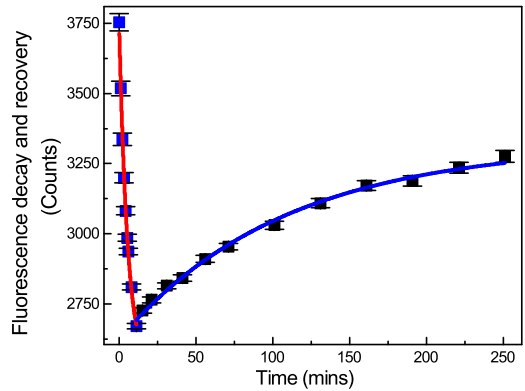}
\end{center}
\caption{Fluorescence decay and recovery of peak amplitude of 1-amino, 2,4-dibromo anthraquinone in PMMA.}
\label{fig:flu-dec-rec}
\end{figure}

We collected the peak amplitudes of the spectra during decay and recovery, which are plotted in the Figure \ref{fig:flu-dec-rec}. The time evolution of the peak amplitude is then fit to the functions:

\begin{equation}
F(t) = F_{0} + F_{d}\exp(-\alpha t) ,
\label{decay-function}
\end{equation}

\noindent and

\begin{equation}
F^{\prime}(t) = F^{\prime}_{0} + F_{r}\exp(-\beta t),
\label{recovery-function}
\end{equation}

\noindent where F(t) is the peak amplitude at time t, $ F_{0} + F_{d} $ and $ F^{\prime}_{0} + F_{r} $ are the amplitude of the spectra at time $ t = 0 $ during decay and during recovery respectively, $ F_{0} $ and $ F^{\prime}_{0} $ are the amplitudes of the spectra at time $ t = \infty$ and $ \alpha $ and $ \beta $ are the decay and recovery rate constants. Figure \ref{fig:flu-dec-rec}, shows that the sample is damaged by about $ 20-30\% $ and recovers to $ 30-50\% $ of its pristine state in 4-6 hours.

\begin{figure}[h!]
\begin{center}
\includegraphics[scale=0.85]{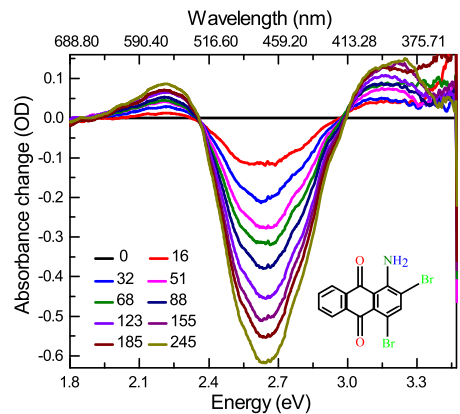}
\end{center}
\caption{Change in the absorption spectrum relative to the pristine state as a function of time.}
\label{fig:change-in-abs}
\end{figure}

A representative change in the absorbance spectra of 1-amino, 2,4-dibromo anthraquinone dye molecule is shown in the Figure \ref{fig:change-in-abs}. The main peak of the absorbance spectra is at 2.653 eV and two satellite peaks at 2.215 eV and 3.144 eV. There are two isobestic points at 2.36 eV and 3.00 eV. The presence of isobestic points in the absorbance spectra indicate that the degraded species of the molecules must pass predominantly through at least one species of photodegradation.

Table \ref{properties_B} shows some of the important properties of dye B at key wavelengths\cite{Prabodh}:

\begin{table}[h!]
\caption{Properties of dye B at key wavelengths (energies). }
\centering
\begin{tabular} {|c|c|c|c|}	\hline \hline
Wavelength 		& comments			& Absorption  		 	& 1/e $ \times $ 				 	\\	
(Energy)		&				& cross section  			& absorption 					\\
 nm (eV) 		& 				& $ (\times 10^{-18} cm^{2}) $	& length ($ \mu $m) 				\\ \hline
488  (2.54 ) 		& pump wavelength 	&					& 							\\
 			& for undamaged 		& 24 					& 88 							\\
 			& molecule 			& 					& 							\\ \hline 	
488 (2.54) 		& pump wavelength 	& 					& 							\\
 			& for damaged  		& 16 					& 140 							\\
 			& molecule 			& 					& 							\\ \hline
575 (2.16) 		& fluorescence peak 	& 					& 							\\
 			& wavelength for 		& 					& 							\\
 			& undamaged   	 	&  					& 	 						\\
 			& molecule 			& 1.2					& 1800 						\\ \hline
575 (2.16) 		& fluorescence peak	& 					& 							\\
 			& wavelength for 		& 					& 							\\
 			& damaged 		 	&  					&  							\\
 			& molecule 			& 2.4 					& 870 							\\ \hline 	526 (2.36)	& isobestic point 1 		& 54 					& 390 							\\ \hline
414 (2.99) 		& isobestic point 2 		& 12 					& 170 							\\ 	\hline
467 (2.653)		& absorption 		& 1.63				& 30    						\\					& peak 	 		& 					& 				\\ \hline \hline

\end{tabular}
\label{properties_B}
\end{table}

\begin{figure}[h!]
\begin{center}
\includegraphics[width = 3.25 in]{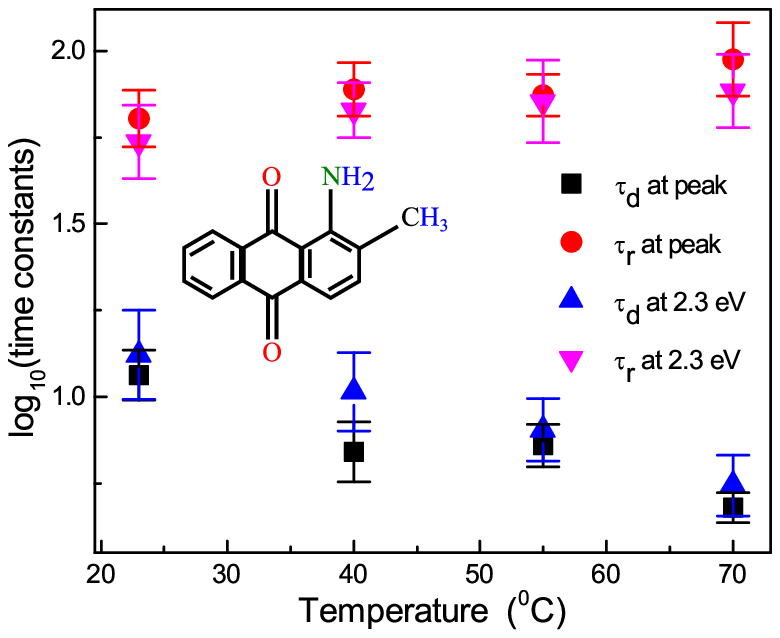}
\includegraphics[width = 3.25 in]{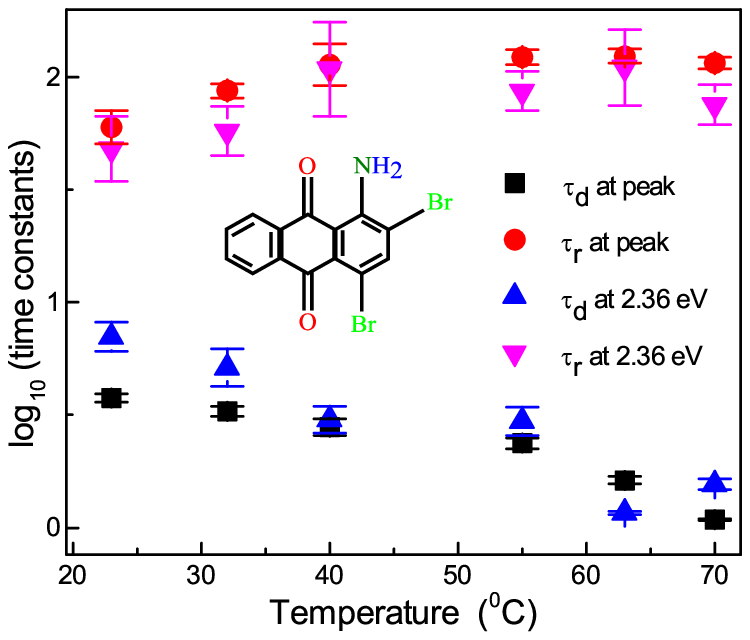}
\end{center}
\caption{Fluorescence decay time constant denoted by $ \tau_d $ and recovery time constant denoted by $ \tau_r $ of dye B (top) and dye A (bottom) as a function of temperature at the peak of the fluorescence spectra and at the isobestic point. The time constant, in both the top and the bottom Figures, is plotted in the natural logarithmic scale.}
\label{fig:Flu-time_constants-B}
\end{figure}

\begin{figure}[h!]
\begin{center}
\includegraphics[scale=0.85]{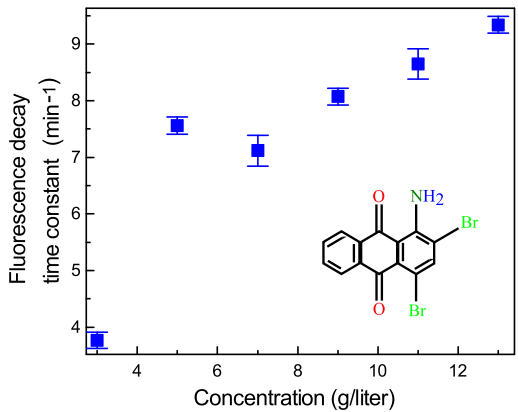}
\end{center}
\caption{Fluorescence decay time constant as a function of concentration.}
\label{fig:conc-dec-time}
\end{figure}

\begin{figure}[h!]
\begin{center}
\includegraphics[scale=0.85]{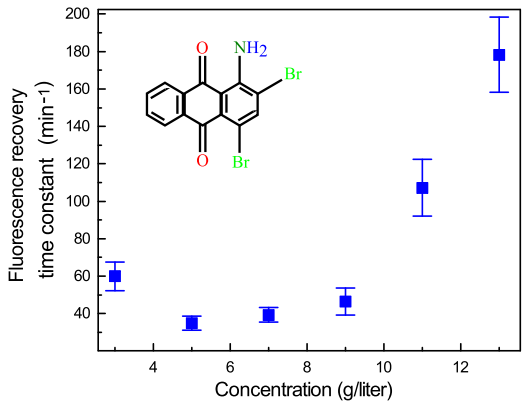}
\end{center}
\caption{Fluorescence recovery time constant as a function of concentration.}
\label{fig:conc-rec-time}
\end{figure}

The temperature and concentration-dependent fluorescence decay and recovery measurements of the dye provides a window into the mechnisms. Figures \ref{fig:Flu-time_constants-B} shows temperature-dependent measurements and Figures \ref{fig:conc-dec-time} and \ref{fig:conc-rec-time} shows concentration-dependent measurements. Each point in Figures \ref{fig:Flu-time_constants-B},  \ref{fig:conc-dec-time} and  \ref{fig:conc-rec-time} is derived from a weighted average fit of many measured data points to account for variations in their uncertainties\cite{Taylor}.

The experimental data for both temperature dependent decay time constant and recovery time constant agree with the domain model, which will be explained later in this section. For comparison, Figure \ref{fig:Flu-time_constants-B} shows the decay and the recovery time constant of DO11. The decay time constant of dye B is much less than that of DO11 and the recovery time constant is larger than that of DO11. Thus, as a function of temperature, dye B is less photo robust as compared to DO11. We attribute this properties to the relatively heavier bromine atom attached to the carbon atoms of the benzene ring of dye B. 

From Figure \ref{fig:conc-rec-time} and similar report showing concentration dependent decay and recovery time constant of DO11\cite{Shiva}, it seems that at the lower concentration, Dye B recovers more quickly than DO11, but as the concentration is increased, the recovery time of Dye B is longer than that of DO11. Hence we can say that at lower concentration, Dye B is more photo robust than DO11, but as the concentration is increased, DO11 is more photo robust than Dye B.  

Whether or not the decay and recovery time constant of dye B as a function of temperature and concentration is greater or less than that of DO11, the self-healing properties of dye B are unique and are mainly due to the heavier bromine atoms attached to the carbon atoms of the benzene ring in place of the methyl group.

The experimental data for concentration-dependent decay time constant is consistent with the domain model; however, the concentration-dependent recovery time constant does not agree with the domain model-not even qualitatively. This will be discussed in detail later in this section.

The environment also plays a crucial role in the decay and recovery rate of the organic dye molecules. Researchers have studied the role of the environment onn the accelerated decay and accelerated recovery of DR1 and found that it may lead to different types of mechanism such as isomerization or photo-oxidation of the molecule in its photodegradation\cite{Stegeman}. Also studied is the role polymer host plays in photodegradation and its recovery in DO11 and found that DO11 shows better recovery from photodegradation in PMMA than when incorporated into a Styrene host matrix\cite{Hung}. Therefore it is beyond doubt that the environment also plays an important role. Even though, we have not carried out any such measurements, we do not rule out any such possibilities. Because of the similarity in the structure of dye B with DO11, we speculate that dye B is also influenced by its environment in phtodegradation and recovery kinetics.

\begin{figure}[h!]
\begin{center}
\includegraphics{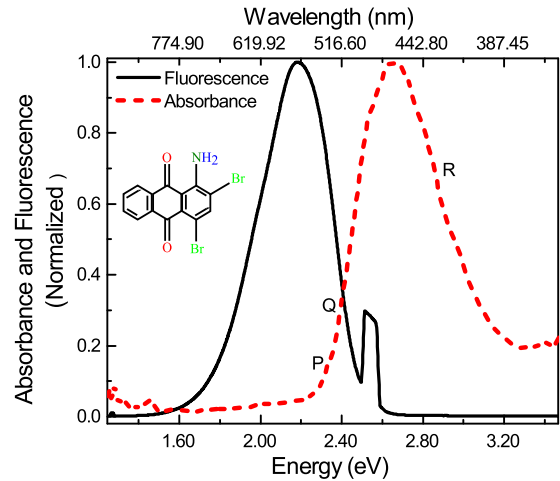}
\end{center}
\caption{Absorbance and fluorescence spectra of 1-amino, 2,4-dibromo anthraquinone. Both spectra are normalized to unity.}
\label{fig:Normalized_abs_flu}
\end{figure}

Next, we estimated the amount of self-absorption of fluorescence signal by the sample before the final signal falls onto the detector. Since the average thickness of the our sample is about 300 $ \mu $m, which is much thicker than 1/e $ \times $ the absorption length of the sample\cite{Prabodh}, self-absorption must be taken into account. A plot of the absorbance and fluorescence spectra as shown in the Figure \ref{fig:Normalized_abs_flu} shows that the two spectra overlap each other in a small region around an energy of 2.40 eV, so self-absorption will be large there.

We consider three points P, Q and R in Figure \ref{fig:Normalized_abs_flu}. Around point P, which lies in the absorbance band, the fluorescence spectrum emitted by the dye is strongly absorbed by the sample itself. Around point Q, which lies in the weak part of the absorbance band, the fluorescence emitted by the dyes is weakly absorbed by the sample. Around point R, which lies in the transparent part of the spectrum, the fluorescence signal is not absorbed. This effect was ignored by previous studies which use ASE as a probe\cite{Embaye, S.K.Ramini, Ramini}.

The Equations that governs decay and recovery are given by \cite{Ramini, Anderson}:

\begin{equation}
\frac{\partial n(z, t)}{\partial t} = N\beta [1 - n(z, t)] -\frac{\alpha}{N} I_{p}(z, t) n(z, t)
\label{population-time}
\end{equation}

\noindent and

\begin{multline}
\frac{\partial I_{p}(z, t)}{\partial z} = -I_{p}(z, t) N [n(z, t) A_{u}(\lambda_{p}) \\
+ \{1 - n(z, t)\} A_{d} (\lambda_{p})],
\label{intensity-time}
\end{multline}

\noindent where n(z, t) is the fraction of undamaged population of molecules, $ I_{p}(z, t) $ is the pump intensity at depth z and time t, $ n^{\prime}(z, t) $ is the population of undamaged molecules at time $ t $ and at depth $ z $, $ \alpha $ is the decay rate per unit intensity, $ \beta $ is the recovery rate, $ A_{u}(\lambda_{p}) $ is the undamaged absorption per unit length at pump wavelength $ \lambda_{p} $, $ A_{d}(\lambda_{p}) $ is the absorbance of the damaged species per unit length at pump wavelength $ \lambda_{p} $, and N is the average number of molecules in a domain.

\begin{figure}[h!]
\begin{center}
\includegraphics[scale=0.85]{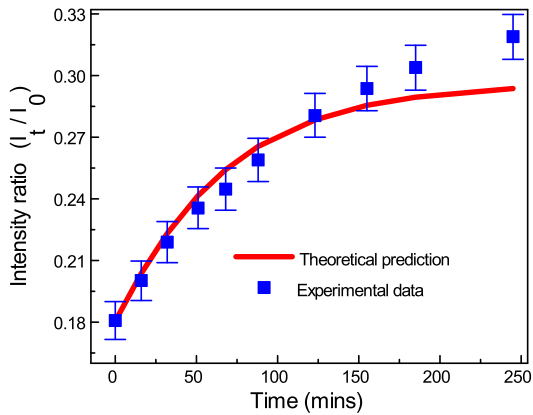}
\end{center}
\caption{The experimental intensity ratio and the theoretical model as a function of time.}
\label{fig:intensity-ratio}
\end{figure}

The programming code was written in Matlab 2013R. For simplicity, we used the Euler method\cite{Epperson, Press, Stan} to obtain the numerical solution of Equations (\ref{population-time}) and (\ref{intensity-time}). We calculated the intensity ratio iteratively\cite{Saad}, with the fitted parameters of the decay rate constant $ \alpha $ and the recovery rate constant $ \beta $, until the theoretical intensity ratio matches with the experimental intensity ratio to a tolerance level of $\epsilon_{max} = 10^{-4} $. The error bar $ \sigma = 2 \times 10^{-3} $ is slightly greater than the tolerance level. The result is shown in the Figure \ref{fig:intensity-ratio}.

\begin{figure}[h!]
\begin{center}
\includegraphics[scale=0.85]{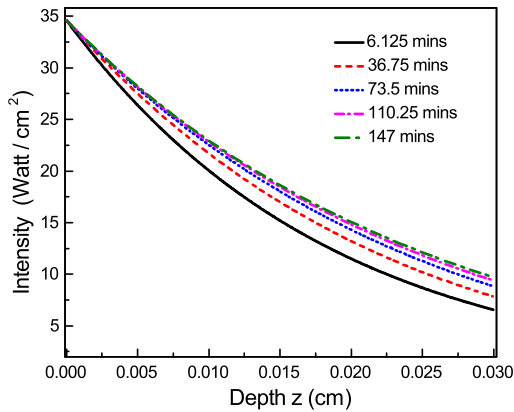}
\end{center}
\caption{The CCrDM's prediction of intensity as a function of time at various depth of the sample.}
\label{fig:intensity-depth}
\end{figure}

\begin{figure}[h!]
\begin{center}
\includegraphics[scale=0.85]{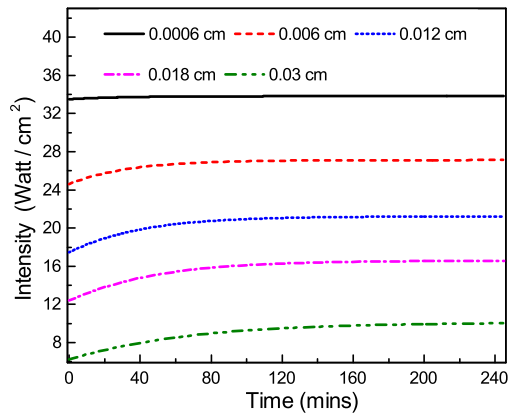}
\end{center}
\caption{The CCrDM's prediction of intensity as a function of depth at various pump exposure times.}
\label{fig:intensity-time}
\end{figure}

After determining the value of the decay constant rate $ \alpha $ and the recovery constant rate $ \beta $, the same code was used  to estimate the intensity and the undamaged molecule fraction as a function of depth and time by using fixed $ \alpha $ and $ \beta $ determined from the fit. Figures \ref{fig:intensity-depth} and \ref{fig:intensity-time} show the model's prediction of the intensity as a function of time of exposure and depth in the sample.

Figure \ref{fig:intensity-depth} shows that the intensity decreases with the depth at all the exposure times. Thus when a particular layer of the sample is not completely damaged, it absorbs some of the pump light so that the next layer will receive less light. As the layer is damaged more and more, it permits a greater amount of light to pass to the next layer. Figure \ref{fig:intensity-depth} shows this feature by the slow change of the intensity at long times. 

\begin{figure}[h!]
\begin{center}
\includegraphics[scale=0.85]{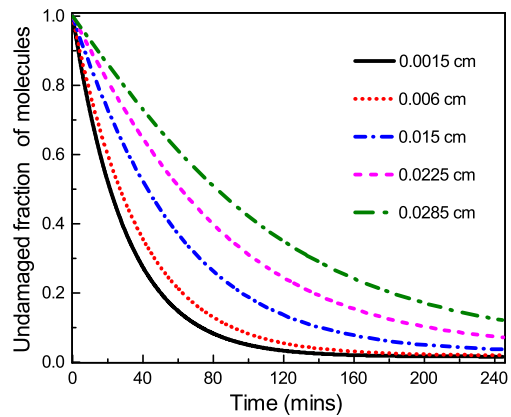}
\end{center}
\caption{CCrDM's estimation of the undamaged molecule fraction as a function of time at various depth.}
\label{fig:n0-time}
\end{figure}

\begin{figure}[h!]
\begin{center}
\includegraphics[scale=0.85]{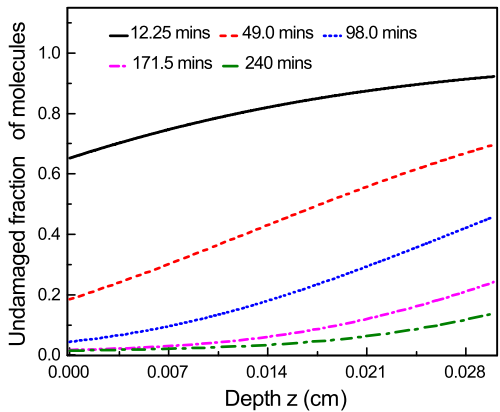}
\end{center}
\caption{CCrDM's estimation of the undamaged molecule fraction as a function of depth at various time.}
\label{fig:n0-depth}
\end{figure}

We simultaneously estimated, using the same code, the evolution of the undamaged molecule fraction as a function of depth and time, as shown in Figures \ref{fig:n0-time} and \ref{fig:n0-depth}. Figure \ref{fig:n0-time} shows that the undamaged molecule fraction decreases quickly at first, then saturates. At longer times the undamaged population is depleted in a layer, so the conversion rates from undamaged species to damaged species tails off. Figure \ref{fig:n0-depth} shows that deeper inside the sample, fewer molecules are damaged because most of the pump light is absorbed by the layers  above it. As a result, the undamaged molecule fraction increases with depth. Figure \ref{fig:n0-depth} also shows that at longer times, the fraction of undamaged molecules decreases since the pump light penetrates deeper inside the sample.

Figures \ref{fig:intensity-depth}, \ref{fig:intensity-time}, \ref{fig:n0-time} and \ref{fig:n0-depth} are consistent with each others and in accordance to our expectation. We can conclude that the more a layer of the dye is damaged, the more it become opaque for the pump laser beam to pass to the next layer. This is seen to be in contrast with DO11 where a layer, after it is damaged, becomes transparent to the next layer\cite{Anderson}.

\begin{figure}[h!]
\begin{center}
\includegraphics[scale=0.85]{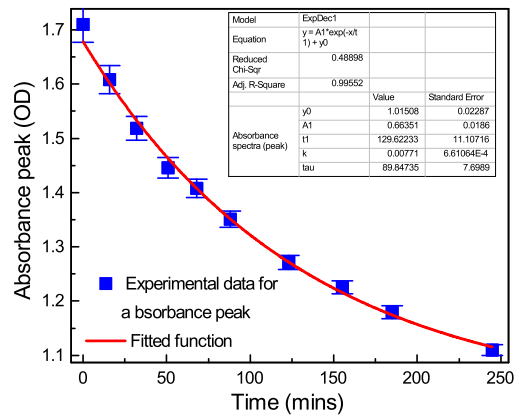}
\end{center}
\caption{Peak of the absorbance spectra as a function of time represented by points and the fit to Equation (\ref{absorbance 1}) represented by the curve.}
\label{fig:abs_peak_time}
\end{figure}

After estimating the amount of self-absorption of fluorescence intensity by the sample before the transmitted signal falls onto the detector, we proceed to find the decay rate constant $ \alpha$ and the recovery rate constant $ \beta $ for a given fluorescence signal and the other parameters of interest in the CCrDM. For this we first need to calculate the time interval to damage half of the pristine molecules. We did this by plotting the peak of the absorbance spectra as a function of time. Figure  \ref{fig:abs_peak_time} shows the plots and the fitted function given by\cite{P.Dhakal}:

\begin{equation}
A_{m}(t, E_{0}) = A_{m}^{(0)} + A_{m}^{(1)} \exp(-t/\tau),
\label{absorbance 1}
\end{equation}

\noindent where, $ A_{m}(t, E_{0}) $ is the measured absorbance at time t and energy $ E_{0} $ and $ A_{m}^{(0)} + A_{m}^{(1)} $ is the amplitude of the absorbance peak at time $ t = 0 $.

The fitting parameters found are shown in the Table \ref{fitting_parameters}.

\begin{table}[h!]
\caption{Fitting parameters obtained from fitting the peak of the absorbance spectra as a function of time to Equation (\ref{absorbance 1}).}
\centering
\begin{tabular} {|c|c|}	\hline \hline
parameters	 				&	value							\\	\hline
						&								\\
$ \tau $ 					& 130 $ \pm $ 11 (mins)					\\	\hline
						&								\\	
$ A_{m}^{(1)} $				& 0.66 $ \pm $ 0.02 (OD)					\\	\hline
						&								\\
$ A_{m}^{(0)} $				& 1.01 $ \pm $ 0.02 (OD)					\\	\hline
\end{tabular}
\label{fitting_parameters}
\end{table}

Using the fitting parameters given in Table \ref{fitting_parameters} and Equation (\ref{absorbance 1}), the time required to damage half of the population of the undamaged species is found to be

\begin{align*}
T_{1/2} = 172.175  \mbox{ mins. } 
\end{align*}

\begin{figure}[h!]
\begin{center}
\includegraphics[scale=0.30]{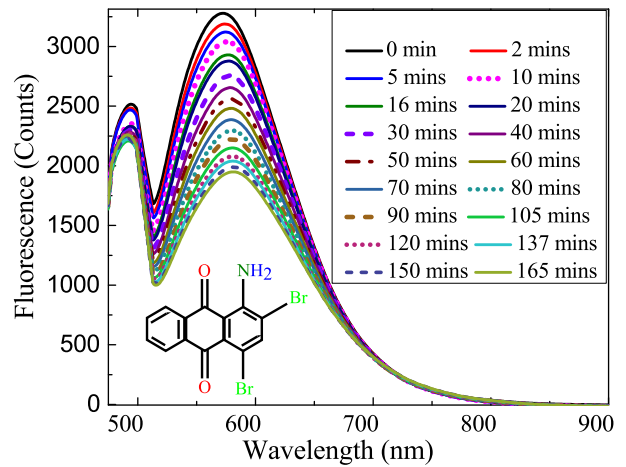}
\end{center}
\caption{Decay of the fluorescence spectrum as a function of time.}
\label{fig:flu-spectra}
\end{figure}

To confirm that at the time $ T_{1/2} = 172.175 $ mins half of the pristine population will be damaged, we also collected the fluorescence spectra while the sample is being damaged by the pump light. The plot is shown in the Figure \ref{fig:flu-spectra}, which shows that at time $ t = 165 $ mins, nearly half of the population is damaged.

Next, we used Equation (\ref{Ad}) to calculate absorbance due to the damaged species\cite{P.Dhakal}:

\begin{equation}
A_{d}(E) = 2 A_{m}(t_{1}, E) - A_{m}(T_{1/2}, E).
\label{Ad}
\end{equation}

\begin{figure}[h!]
\begin{center}
\includegraphics[scale=1.00]{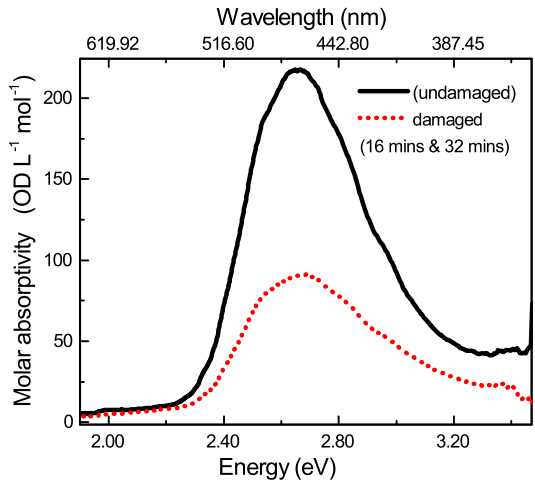}
\end{center}
\caption{Spectra due to the damaged species. For comparison, the undamaged spectrum is also shown.}
\label{fig:Ad_Au}
\end{figure}

/noindentation The result is shown in the Figure \ref{fig:Ad_Au}. For comparison, we have also shown the absorbance due to the undamaged sample. The peak of the damaged spectra is also about half of the peak of the undamaged spectra, as we expected. 

Next we found the decay rate constant $ \alpha $ and recovery rate constant $ \beta $ for the fluorescence signal and calculate the CCrDM parameters. Our data for both temperature dependent decay time constant and recovery time constant agrees with the domain model. The decay and the recovery behavior of these data as a function of temperature can be understood as follows. The dye molecules form domains mediated by the polymer chain. As the temperature increases, the average domain size decreases because the thermal energy gained by the domain breaks it apart into the smaller domains. However, there are many domains of varying sizes within the sample. In a particular domain where some of the molecules have already undergone photo decay, an additional exposure to the pump beam causes a greater photo decay than in a domain of the same size but with no damaged molecules. Hence, the decay time constant decreases as a function of temperature.

Similarly, at higher temperature the domains are smaller with more damaged species. Hence, the damaged species takes more time to recover to their original state. At lower temperature, the domains are relatively larger. As a result, the damaged species recover more  quickly. Therefore, the data shown in Figures \ref{fig:Flu-time_constants-B} are consistent with the domain model. 

Similarly, the higher the concentration of the dye molecules in the host polymer matrix, the larger the domain size. As a result, the sample takes more time to decay. The decay time constant as a function of concentration is also consistent with the domain model. There is anomalous behavior at the concentration of 5 g/liter. The behavior at this concentration is not fully understood and has also been observed by others in similarly structured DO11 dye\cite{}. Further study is required to confirm this result and understand its cause. Hence aside from the 5 g/l value, the behavior of the decay time constant of the molecules as a function of concentration shown in Figures \ref{fig:conc-dec-time} is consistent with the domain model.

Our data for the recovery time constant as a function of concentration contradicts the domain model. According to the domain model, the higher the dye concentration, the greater the undamaged population. Hence, it should recover quickly. The data in Figure \ref {fig:conc-rec-time} shows the opposite behavior. The reason for this anomalous behavior is not well understood yet. There could be other mechanisms at work, which leads to the recovery of the damaged species. This requires more study.

The details of how self-absorption is taken into account for our geometry is given by the reference\cite{P.Dhakal}.

\begin{figure}[h!]
\begin{center}
\includegraphics[scale=0.85]{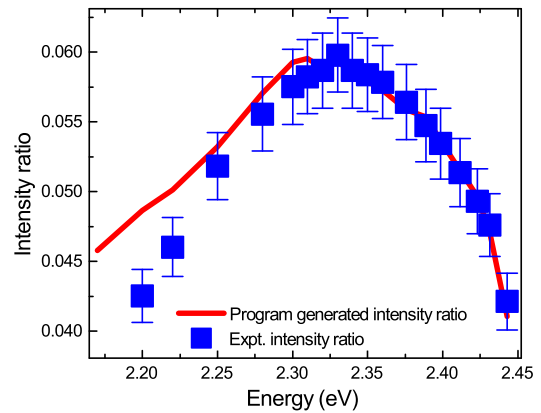}
\end{center}
\caption{The experimental fluorescence intensity ratio and the theoretical model.}
\label{fig:intensity-ratio-2}
\end{figure}

The theoretical prediction is compared with the experimental data as shown in Figure \ref{fig:intensity-ratio-2}. The two agree within the limits of the error bars except for 2-3 points at the lower energy limit.

After fitting the theoretical prediction with the experimental data, we found the decay rate constant $ \alpha $ and recovery rate constant $ \beta $ as the fitting parameters. The critical intensity denoted by $ I_{c} $ is the ratio of $ \beta $ to $ \alpha $  It measures the intensity of the pump beam at which no degradation is observed. Materials with large $ I_{c} $ are desired because the higher the value of the critical intensity, the more photo robust the material is to degradation. Our data gives the critical intensity of $ I_{c} = (1.10 \pm 0.45) \times 10^{-5} (W cm^{-2}) $, which is quite high.

The results for the rate constants are summarized in the Table \ref{alpha, beta}. For comparison, the corresponding rate constants for DO11 is also listed in Table \ref{alpha, beta}.

\begin{table}[h!]
\caption{The decay and recovery rate constants for fluorescence self absorption in 1-amino, 2,4-dibromo anthraquinone and DO11. }
\centering
\begin{tabular} {|c|c|c|c|}	\hline \hline
Molecules	   & Decay constant             	      & Recovery            	   	 & $ I_{c} $		        		\\	
               	   & $ \alpha $                           	      & rate                        	    	 &                       		           \\
		   & $ cm^{2}/(W\times min $)           & $ \beta$ ($ min^{-1} $)  	 &	(W/$ cm^{2} ) $	    	 \\
               	   & $ \times 10^{-2} $		      &	$ \times 10^{-6} $	        	 & $ \times 10^{-6} $    		 \\ 	\hline
		   & 					      & 			                	 & 	                    			 \\ 	
1-amino, 2,4-  & $ 11.7 \pm 2.4  $  		      & $ 1.30 \pm 0.72 $ 		 &	$ 11.0 \pm 4.5 $ 		 \\ 		
dibromo AQ     &					      &					 &				            \\	\hline
DO11\cite{Anderson}& $ 15.5 \pm 3.9 $	      & $ 22.8 \pm 9.5 $            	 &  $ 150 \pm 51 $  			 \\	\hline
		   &					      &					 & 	           	        		 \\	
DO11\cite{Ramini} & 12.2 $ \pm $ 0.2		      & $ 1.28 \pm 0.12 $             	 &    $ 10.5 \pm 1.2 $        	 \\	\hline

\end{tabular}
\label{alpha, beta}
\end{table}

The numerical value of the decay rate constant $ \alpha $ we have calculated for 1-amino, 2,4-dibromo anthraquinone agrees with that for DO11 obtained by Anderson et al within the limit of experimental uncertainty, but does not agree with the same quantity obtained by Ramini et. al. On the other hand, the recovery rate constant $ \beta $ does not agree with that calculated by both Anderson et al. and Ramini et al. The reason for this discrepancies is that the nature of 1-amino, 2,4-dibromo anthraquinone is  different from DO11, even though both of these anthraquinone derivatives have similar structure. Furthermore, the three separate studies employed different techniques to determine the population of undamaged molecules. As a result, this might have led to at least some of the differences.

Upon obtaining the value of the decay rate constant $ \alpha $ and the recovery rate constant $ \beta $, we we tested our data to the two population correlated chromophore domain model proposed by Ramini et al\cite{Ramini, S.K.Ramini}. Using the depth effect given by Equations (\ref{population-time}) and (\ref{intensity-time}), we calculated the number density of molecules $ \rho $, average number of molecules in a domain N and partition function for a single molecule in a domain z as the fitting parameters. From the partition function for a single molecule in a domain z, we calculated the free energy advantage per molecule $ \delta\mu $, which is the energy required to extract one molecule out of the domain. The results are listed in the table \ref{ccrdmodel-value}. For comparison, we have also listed the value of the domain model parameters for DO11 obtained by Ramini et al and Anderson et al\cite{Ramini, Anderson}.

\begin{table}[h!]
\caption{Parameters from correlated chromophore domain model.}
\centering
\begin{tabular} {|c|c|c|}	\hline \hline
parameters		&	value for			& value for									\\	
			& 1-amino, 2,4-dibromo 		& DO11									\\	
			& anthraquinone			& 										\\	\hline
			&					&										\\	
N			& 61 $ \pm $ 4			& 40 $ \pm $	5\cite{S.K.Ramini}						\\	\hline
			&					&										\\
$\rho$		& 0.050 $ \pm $ 0.003		& 0.02\cite{Ramini}, 0.019\cite{Anderson}				\\	\hline
			&					&										\\	
$\delta\mu$ (eV)	& 0.53 $ \pm	$ 0.06			& 0.29 (eV)$\pm$ 0.01\cite{Ramini, Anderson}			\\	\hline
\end{tabular}
\label{ccrdmodel-value}
\end{table}

From the comparison given in the Table \ref{ccrdmodel-value}, we can see that the numerical value of CCrDM parameters for dye B are higher than the corresponding value for DO11. The difference is not unexpected because the atoms or group of atoms attached to carbon atoms of the benzene ring is different in two dye molecules. Also the atoms are attached to different carbon atoms of the benzene rings. Equally important is the fact that both Ramini et al did not take into account of the depth effect. The time scale of the measurement in their study is also different. The time scale  for the recovery in our study is about 5-6 hours, during which we did not observe the irreversibly damaged species. Anderson et al. has observed the irreversibly decayed species after taking the recovery measurements for about 20-24 hours\cite{Anderson}, which is a much longer amount of time as compared to our measurements.

\section{Conclusion}

Using fluorescence spectroscopy, we studied photodegradation and recovery as a function of temperature and concentration of 2,4-dibromo anthraquinone molecule doped in host PMMA polymer. Based on our observation, we compared the recovery rate of 2,4-dibormo anthraquinone molecule with DO11. We found that it takes more time to damage DO11 than 2,4-dibormo anthraquinone dye and it takes less amount of DO11 to self-heal than  2,4-dibormo anthraquinone dye molecules. In other word, we found that DO11 is more photo robust than the 2,4-dibormo anthraquinone dye molecule. We also found that the surrounding environment as well as the host polymer matrix also plays a crucial role in the recovery of the dye molecules. 

The absorption spectra due to the damaged species of 2,4-dibormo anthraquinone dye molecule was determined. Using the absorption due to the damaged species, we also accounted for self absorption. We estimated the intensity and population of undamaged molecule fraction as a function of depth of the sample and the time of exposure of the pump light. Our study showed that the more a particular layer of the 2,4-dibormo anthraquinone dye molecule is damaged, the more it become opaque to the incident pump light to enter the next layer. This finding is in contrast to DO11 where the hyperbolic tangent types of curve showed that, upon damaged by the pump light, the layer becomes more transparent to the pump light. Finally we  tested the Chromophore Correlated Domain Model (CCrDM) and calculated the model parameters as well as the fluorescence decay and recovery rate constants of the dye, which characterize the decay and the recovery rates of the dye. Our data qualitatively agrees with the model.

There are many questions that remain unanswered. In other to understand the self-healing properties of 2,4-dibormo anthraquinone dye molecule, further independent studies are required that explore the effect of pump beam intensity, the effect of the gaseous environment in which accelerated recovery may take place, the effect of electric field in the recovery of the dye molecule etc. Furthermore, our data does not reveal any information about the real physical existence of domains. Direct X-ray scattering measurements and Fourier Transform Infra Red (FTIR) measurements may reveal important aspects of the presence of the domain and their size distribution, the nature of the domain and the bonds holding them together. Even though our data qualitatively agrees with the CCrDM, our results alone should not be understood as the underlying mechanism for self recovery of the anthraquinone dye molecules. The other mechanism or a combination of the other mechanism such as photo-induced chemical reaction \cite{}, more species involved\cite{Ben_Anderson, B._Anderson} and the composition of the host polymer matrix\cite{Hung} may also contribute the self recovery of anthraquinone types of molecules.

\section{Acknowledgments}

We would like to thank Wright Patterson Air Force Base and Air Force Office of Scientific Research (FA9550- 10-1-0286) for supporting this research.

\bibliographystyle{osajnl2}
\bibliography{\bibs}

\end{document}